%% file: limitsv7.tex
\documentclass[aps,prd,twocolumn,groupedaddress,nofootinbib]{revtex4-2}
\usepackage{tensor}
\usepackage{amsmath}
\usepackage{amssymb}
\usepackage{mathtools}
\usepackage{graphicx}
\usepackage{subfig}
\usepackage{upgreek}
\usepackage{wrapfig}
\usepackage{float}
\usepackage[
	colorlinks=true,
	linkcolor=blue,
	citecolor=blue,
	filecolor=magenta,
	urlcolor=blue
]{hyperref}

\newcommand{\be}{\begin{equation}}
\newcommand{\ee}{\end{equation}}
\newcommand{\bea}{\begin{eqnarray}}
\newcommand{\eea}{\end{eqnarray}}
\newcommand{\Lm}{\mathcal{L}_{m}}

\begin{document}
\title{Nuclear Limits on Non-Minimally Coupled Gravity}
\author{Sarah B. Fisher}
\author{Eric D. Carlson}
\email{ecarlson@wfu.edu}
\affiliation{Department of Physics, Wake Forest University, 1834 Wake Forest Road, Winston-Salem, North Carolina 27109, USA}
\date{\today}

\begin{abstract}
We explore alternate theories of gravity where the gravitational term in the Lagrangian $\frac{R}{8\pi G}$ is replaced by a function $f_1(R)$ and the matter Lagrangian is multiplied by a function $f_2(R)$.  We argue that nuclear physics can provide strong experimental constraints on such theories.  In particular using energy conditions on the pressure in the $^4$He nucleus, for $f_1(R)=\frac{R}{8\pi G}$ and $f_2(R)=1+\lambda R$, we find a limit of $|\lambda| < 5\times 10^{-12}\, \hbox{m}^2$, more than thirty orders of magnitude stronger than the previous limit.

\end{abstract}

\maketitle

\section{Introduction}
Although Einstein's theory of general relativity is very successful, alternative theories have received a great deal of attention.  For example, in $f(R)$ gravity, the gravitational action is an arbitrary function of the scalar curvature $R$.  Nojiri and Odintsov \cite{Nojiri_2004} proposed an alternative where the dark energy Lagrangian is instead multiplied by a function of the curvature.  Bertolami {\it et al.} \cite{Bertolami_2007} proposed multiplying the entire matter Lagrangian by this function of the curvature and combining it with the non-trivial gravitational action, and this combination was subsequently used to reproduce the observed acceleration of the universe \cite{Bertolami_2010a, Bertolami_2014}, galactic rotation curves \cite{Bertolami_2010b, Harko_2010,Bertolami_2010c}, the dynamics of galaxy clusters \cite{Bertolami_2012}, and inflation in the early universe \cite{PhysRevD.83.044010}.

While a number of people have explored such theories, relatively few have considered the constraints imposed by pre-existing data \cite{Bertolami_2008b,Bertolami_2013,Castel_Branco_2014,March_2019}.  In addition, research focuses overwhelmingly on astrophysical scenarios.  While this makes some sense, as the initial motivation  came from this area, there is no reason to expect that the area in which one hopes to see benefits is the best place to look for constraints.  March {\it et al.} studied the effects of non-standard gravity on ocean experiments \cite{March_2019} and found effects considerably stronger than such astrophysical limits.  We argue that this result is not surprising, because the strongest constraints come from  large density gradients, which can be measured more readily in terrestrial than astrophysical observations.  We find that looking at nuclei can produce limits many orders of magnitude better than previous limits.

The paper is organized as follows.  In Sec.~\ref{The Lagrangian} we discuss the appropriate choice of Lagrangian to describe a perfect fluid, and contend that the appropriate on-shell matter Lagrangian is given by $\Lm = - \rho$.  In Sec.~\ref{The Test System} we analyze what systems can be studied to provide constraints on non-standard gravity coming from $f_2(R)$, and conclude that nuclear physics is a good choice, because of the large density gradient.  In Sec.~\ref{Example} we apply this to a simple model where $f_2(R) = 1 + \lambda R$, and derive limits on $\lambda$. In Sec.~\ref{Energy Conditions} we discuss how energy conditions can be used to obtain limits and apply it to our model.  In Sec.~\ref{Conclusions} we summarize our conclusions and compare to previous work. Throughout our paper, our conventions are $c=1$, metric signature $(+---)$, and curvature $R_{\mu\nu}=R_{\mu\nu}^{MTW}$, and $R=-R_{MTW}$ where $MTW$ refers to the conventions of Misner, Thorne, and Wheeler~\cite{Misner1973}.

\section{The Lagrangian}\label{The Lagrangian}
For concreteness, we will consider a form of the action $I=\int d^4x\sqrt{-g}{\cal L}$ which is prevalent in the literature, 
\be\label{action}
{\cal L}=-\frac{1}{2}f_1(R)+ f_2(R)\Lm \; ,
\ee
where $\Lm$ is the matter Lagrangian.  With this action, the gravitational field equations are given by
\be\label{general EOM}
-2\frac{\partial{\cal L}}{\partial R} R^\mu_{\;\;\nu}-\frac{1}{2}f_1(R)g^\mu_{\;\;\nu} +2(\nabla^\mu\nabla_\nu-g^\mu_{\;\;\nu}\Box)\frac{\partial{\cal L}}{\partial R}  = f_2(R)T^\mu_{\;\;\nu}\; ,
\ee
where $\Box = \nabla^\mu\nabla_\mu$ is the d'Alambert operator and the stress-energy tensor is given as usual by
\begin{equation}\label{StressEnergy}
T^{\mu\nu}=- \frac{2}{\sqrt{-g}}\frac{\delta}{\delta g_{\mu\nu}} \left(\sqrt{-g}\Lm\right) \; .
\end{equation}
It should be noted that $T^{\mu\nu}$ is derived from the matter Lagrangian, and, in the presence of non-minimal gravity coupling, does not necessarily correspond to what one would physically measure  \cite{Fisher_2019}.  Ordinary gravity is recovered by setting $f_1(R) = \frac{1}{8\pi G}R$ and $f_2(R)=1$.  In this class of theories, stress-energy is not generally conserved; instead, we have
\be\label{conservation}
\nabla_\mu\left[f_2(R)\ T^\mu_{\;\;\;\nu}\right]=-\Lm\nabla_\nu f_2(R) \; .
\ee

As is common, we will consider the case where the matter can be described as a perfect fluid, so that 
\be T_{\mu\nu} = (\rho+p)u_\mu u_\nu - p g_{\mu\nu} \; , \ee
where $\rho$ is the comoving energy density, $p$ the pressure, and $u_\mu$ the four-velocity satisfying $u_\mu u^\mu = 1$.  This density and pressure should in principle be derivable directly from the matter Lagrangian $\Lm$, and should be functions of only the number density $n$ and entropy per particle $s$.  This requires care, as pointed out by \cite{Bertolami_2008}.  In ordinary gravity there are various options for the perfect fluid Lagrangian which yield identical equations of motion, but they are not the same in the presence of non-trivial $f_2(R)$ \cite{Faraoni_2009} because the matter Lagrangian appears in Eq.~(\ref{general EOM}) in its own right.

To resolve the issue, one must look more closely at how the Lagrangian for a perfect fluid works.  It is common to simply assume a form such as $\Lm = - \rho$ or $\Lm=p$.   These Lagrangians are normally derived in standard gravity by imposing equations of motion on a more detailed underlying Lagrangian.  Therefore, they cannot be used to derive the equations of motion.  They do not even yield the correct stress-energy tensor.   Following \cite{Brown_1993}, the form $\Lm = -\rho$ can be derived from
\be\label{Lagrangian}
\Lm= -\rho\left(n,s\right)+\beta_{\!A}J^\mu\nabla\!_\mu\alpha^A+\theta\nabla\!_\mu(sJ^\mu)+\phi\nabla\!_\mu J^\mu \; ,
\ee
and the form $\Lm = p$ can be derived from
\be
 \Lm= -\rho\left(n,s\right)+J^\mu\left(\beta_{\!A}\nabla\!_\mu\alpha^A-s\nabla\!_\mu\theta-\nabla\!_\mu\phi\right) \; ,
\ee
where $J^\mu = n u^\mu$,\footnote{Note that our $J^\mu$ differs from \cite{Brown_1993} by a factor of $\sqrt{-g}$ because we define it to be a vector, rather than a vector density.} $\alpha_A$ ($A=1,2,3$) are fluid coordinates that label each comoving component of the fluid, and $\beta_A$, $\theta$, and $\phi$ are Lagrange multiplier fields.  We consider $J^\mu$ as an independent variable, with $n$ and $u^\mu$ derived from $J^\mu$.  These Lagrangians are equivalent in ordinary gravity because one can transform between them by integrating by parts.  

In non-minimally coupled gravity the equivalence breaks down because the matter Lagrangian is no longer a full term of the action, and therefore integrating it by parts is no longer mathematically valid.  If one derives the matter equations of motion from the different forms of the Lagrangian, one ends up with factors of $f_2(R)$ in different places, sometimes inside derivatives, depending on which equations require integration by parts.  This is in conflict with the claim of \cite{Bertolami_2008} that the equations of motion remain unchanged.

Using an inappropriate Lagrangian can yield non-conservation of particle number, as pointed out for barotropic fluids by \cite{PhysRevD.88.027506}.  The correct form must yield the conservation laws the Lagrange multipliers were introduced to yield, conserving particle number $\nabla_\mu J^\mu = 0$, comoving entropy density $\nabla_\mu (s J^\mu)$, and comoving fluid coordinates $J^\mu \nabla_\mu \alpha_A = 0$, as pointed out by \cite{Fisher_2019}.  This can be done by keeping the factor of $f_2(R)$ attached to the Lagrange multipliers, as is done in Eq.~(\ref{Lagrangian}), yielding $\Lm=-\rho$.  This conclusion matches \cite{Minazzoli_2012}.  In fact, $f_2(R)$ can be absorbed into the Lagrange multipliers, yielding the form of the Lagrangian that \cite{Bertolami_2008} eventually declares to be correct.  We therefore agree with their final result.

\section{The Test System}\label{The Test System}
In order to keep the theory physically viable, we require that the modifications be perturbative.  That is, we require that $f_1(R)-\frac{R}{8\pi G}$ and $f_2(R)-1$ be sub-leading order under ordinary circumstances.

Under such conditions, the change to the matter equations of motion, found in \cite{Bertolami_2008}, will also be sub-leading order.  The gravitational equations of motion are a different matter.  In particular the term $2(\nabla^\mu\nabla_\nu-g^\mu_{\;\;\nu}\Box)\frac{\partial{\cal L}}{\partial R}$ is potentially of different order from the other changes.  The derivatives in this term can act on either $R$ or $\rho$ and will tend to have the physical effect of smoothing out the quantity being differentiated.  A smoothing of $R$ would quite possibly go unnoticed, as it amounts to a change in gravity at small scales and it is difficult to measure gravity on small scales.  A smoothing of $\rho$ is a different matter.  We can often measure density very precisely, and it is not at all difficult to find sharp density gradients---as long as one is not restricting one's attention to astrophysical scenarios.  We therefore expect the strongest limits on theories with non-trivial $f_2(R)$ to be found by studying systems with strong density gradients.  The strongest known density gradients are found in nuclear physics.  We will look at the alpha particle, also known as the $^4$He nucleus, as we feel it provides the best combination of high density gradient, quality measurements, and spherical symmetry.

We will not take a first-principles approach to the $^4$He nucleus, because nuclear structure can be quite complex and difficult to predict from underlying physics.  A rigorous calculation of exactly how strongly such theories are constrained might require this, as the presence of new interactions in modified gravity might well significantly change the structure of a nucleus, but our concern is to illustrate the principle, not to map out the precise limits imposed on various theories.  We will therefore consider the approximation that a nucleus can be described as a perfect fluid to be adequate for our purposes.

For the same reason, we will not try to solve for the density self-consistently.  Instead, we will assume the energy density is proportional to the nucleon density.  Since $^4$He is an isospin singlet, we anticipate that the mass density will be well-approximated by nucleon mass times twice the proton density, which in turn is traced by the charge density. The charge density will presumably be determined exclusively from the standard model matter Lagrangian, and hence not modified by the presence of non-minimal gravitational couplings.  The perfect fluid Lagrangian $\Lm$ is also rooted in the standard model Lagrangian, and hence we would expect the energy density coming from it, rather than the physically measured density, to be proportional to the charge density.  

For the charge density, we will use the ``experimental" charge distribution from \cite{Antonov_2005} (quotes are theirs).  This distribution appears to be calculated by ``model independent" techniques rather than actually measured, and it has no error bars, but they also give the form factors calculated from this distribution and the experimentally measured form factors.  The two seem to compare well, although they are not on the same graph.  A rigorous calculation would require taking the error bars on the measured form factors, using them to calculate a range of possible charge distributions, and comparing these to nuclear structure calculations incorporating modified gravity.

In order to ensure that the lack of rigor in our calculations does not produce spurious constraints, we will use two models for the nuclear density and compare our results.  We fit the charge distribution graphed in \cite{Antonov_2005} with a function of the form
\be\label{density} \rho_e = N e^{ - \left[(ar)^4+b^4\right]^{1/4}} \; , \ee
where $a=2.677\,{\rm fm}^{-1}$ and $b=3.549$.  Then $N = 8.00\,{\rm u/fm}^3$ yields the correct total mass.  We will also consider a simple Gaussian,
\be\label{Gaussian density} \rho_g = \rho_0 e^{-\frac{r^2}{2\sigma^2}} \; , \ee
where $\rho_0=0.278\,{\rm u/fm}^3$ and $\sigma=.9705\hbox{ fm}$, which is fitted to the measured mass and rms radius $r_{\rm rms} = \sqrt3 \sigma = 1.681\,{\rm fm}$ \cite{Sick_2015}, but clearly of a different shape.  It should be noted that any distribution that has a cusp in the density distribution will yield nonsense results due to the contributions from $\Box \rho$, and therefore commonly-used distributions like $\rho = \rho_0\{1 - \exp[(r-R_{1/2})/a]\}^{-1}$ will not work.  We will then find constraints on modified gravity by calculating the pressure needed to maintain each shape, and noting the threshold at which it exceeds various energy conditions.

\section{The Example Model}\label{Example}
To illustrate the value of looking to nuclear structure for constraints, we will apply the approach to the simplest model with non-minimal coupling.  This model is given by
\begin{subequations}
\bea
f_1(R)=\frac{1}{8\pi G}R \; ,\\
f_2(R)=1+\lambda R \; .
\eea
\end{subequations}
The only prior constraint on this model that we can find in the literature is $|\lambda| \ll 1.65\times 10^{20}\,{\rm m^2}$, found by \cite{Bertolami_2008b} by assuming that changes to the Sun's density profile must be perturbative.

With this choice, the gravitational equation of motion becomes
\bea\label{specific EOM}
&&8\pi G (1 + \lambda R)T^\mu_{\;\;\nu} = (1+16 \pi G \lambda \rho)\left(R^\mu_{\;\;\nu} - \frac12 R g^\mu_{\;\;\nu} \right) \nonumber \\
&&\qquad\qquad{} - 16\pi G \lambda \left(\nabla^\mu \nabla_\nu - g^\mu_{\;\;\nu}\Box - \frac12 R g^\mu_{\;\;\nu}  \right)\rho \; ,
\eea
and its divergence becomes
\be\label{conservation2} \nabla_\mu\left[(1+\lambda R)T^\mu_{\;\;\;\nu}\right]=\lambda\rho\nabla_\nu R \;.  \ee
We write the metric in the form
\be\label{metric} ds^2=e^{\nu(r)}dt^2-h(r)dr^2-r^2d\Omega^2 \; . \end{equation}

We wish to solve Eqs.~(\ref{specific EOM}) and (\ref{conservation2}), but solving them exactly generates complexities that yield little insight.  We therefore seek approximations that are appropriate for the range in which we expect to find limits. If $| \lambda | \lesssim r_n^2$, where $r_n$ is the characteristic size of the nucleus, the effects of the new term are perturbations on the unmeasurably small ordinary gravitational attraction of a nucleus to itself.  On the other hand, if $8 \pi G | \lambda | \rho \gtrsim 1$, then the modified gravity would cause the metric to have large deviations from flat space, possibly even a singularity.  Since such a large gravitational effect in a nucleus should cause substantial changes in scattering experiments, we exclude the region $8 \pi G | \lambda | \rho \gtrsim 1$ as not physically plausible.  For our computations, we therefore assume that $| \lambda| \gg r_n^2$ and $8 \pi G | \lambda | \rho\ll 1$, which numerically means we are assuming $10^{-30}\,{\rm m^2} \ll | \lambda | \ll 10^8\,{\rm m}^2$.  We will also assume that the pressure is not much greater than the mass density, so $|p| \lesssim \rho$.  When encountering derivatives on the density, we will treat derivatives as being of order $r_n^{-1}$, so that, for example, $ \rho^{\prime\prime} \sim r_n^{-2} \rho$, where primes denote derivatives with respect to $r$.

We begin by looking at the trace of Eq.~(\ref{specific EOM}), which is
\be\label{trace} 48\pi G \lambda \Box \rho - 8 \pi G (\rho-3p) = R -8 \pi G \lambda R(\rho +3 p) \; . \ee
Making use of our approximations $|\lambda \rho^{\prime\prime}| \gg \rho$ and $8\pi G |\lambda| \rho \ll 1$, this simplifies to
\be\label{Reqn} R = 48 \pi G \lambda \Box \rho \; . \ee
Note that in the nuclear region the contribution to the curvature from new physics is larger than that from conventional gravity.

Before writing out the d'Alambertian explicitly, we need to get some idea of the size of the metric deviations from flat spacetime. We find these by looking at the temporal and radial components of Eq.~(\ref{specific EOM}). The temporal component can be written as
\bea\label{heq}
(1+16 \pi G \lambda \rho) \frac{d}{dr}\!\!\left(r h^{-1} -r\right) &=&   16 \pi G \lambda  h^{-1/2} \frac{d}{dr}\!\!\left(\rho^\prime r^2 h^{-1/2} \right) \nonumber \\
&&{}+8\pi G \rho r^2  .
\eea
Our approximations mean that the first factor on the left is just one and we can neglect the final term on the right. Assuming the metric perturbation is small, $| h-1 | \ll 1$, we can bring the $h^{-1/2}$ inside the derivative to leading order, and integrate to yield
\be\label{heq2} h= 1 - 16\pi G \lambda \rho^\prime r \; , \ee
where the constant of integration is chosen so that $h$ is non-singular at the origin. 

The difference between the temporal and radial components of Eq.~(\ref{specific EOM}) can be put in the form
\bea\label{fStart}
-\left[1 + 8\pi G \lambda(2 \rho - r \rho')\right] \left(\frac{h'}{h}+\nu'\right) =16\pi G \lambda r \rho''&&   \nonumber \\
{}+8\pi G h r (1 + \lambda R) (\rho +p) \quad &&\eea
If we substitute Eq.~(\ref{Reqn}) into (\ref{fStart}), it is not hard to see that the first term on the right dominates the others, and the factor on the left is just one. Substituing  (\ref{heq2}) we have
\be\label{feq1} \nu' = 16\pi G \lambda \rho' \; .\ee
Note that this implies $r_n \nu' \ll 1$.  We see from Eqs.~(\ref{heq2}) and (\ref{feq1}) that the condition that the metric be perturbative requires our approximation $8\pi G |\lambda| \rho \ll 1$, as asserted previously.  We also note that the deviation of metric from flat spacetime from new physics exceeds the contribution from standard gravity.

Expanding the d'Alambertian explicitly in terms of metric components and applying our approximations, Eq.~(\ref{Reqn}) becomes simply
\be\label{Reqn2} R = -48 \pi G \lambda \left(\rho'' + \frac{2}{r} \rho'\right) \; . \ee
We then proceed to find the pressure, starting from Eq.~(\ref{conservation2}), which is
\be\label{PDerivative}
\frac{d}{dr}\left[(1+\lambda R) p\right] = -\frac12 \nu'(\rho+p)(1+\lambda R) - \lambda \rho R'\; .\ee
In standard gravity with $\lambda=0$, Eq.~(\ref{PDerivative}) would lead to the Tolman-Oppenheimer-Volkoff equation, but by substituting Eqs.~(\ref{feq1}) and (\ref{Reqn2}) we find that, in our case, the last term on the right dominates.  Dropping the other terms, integrating by parts, and then substituting Eq.~(\ref{Reqn2}), we find
\be\label{P} p =\frac{48\pi G \lambda^2 \left(\rho\rho'' + \frac{2}{r}\rho\rho' -\frac12 \rho^{\prime 2}+2\int_r^\infty \frac{1}{r}\rho^{\prime 2} dr\right)}{1-48\pi G \lambda^2 \left(\rho'' +\frac{2}{r}\rho' \right)} \ee  

\section{Energy Conditions and Results}\label{Energy Conditions}
It should be noted that Eq.~(\ref{P}) was derived without reference to any equation of state, and therefore does not describe the pressure produced by a given mass distribution.  It describes the pressure {\it required} to maintain a given mass distribution in the presence of the coupling we have chosen to examine.  A rigorous calculation of the sort discussed in Sec.~\ref{The Test System} would have us also calculate the pressure from the equation of state, and require that the two calculations align.  Because nuclear equations of state are hard to calculate, we will instead consider plausible constraints an equation of state might satisfy, in the form of energy conditions.

Energy conditions, like so many other things, require care when working in the context of modified gravity.  As discussed in \cite{Curiel_2017}, there are various ways of expressing energy conditions, which are equivalent under the assumption of the Einstein field equation.  In modified gravity the Einstein field equation does not hold, and the equivalence breaks down.  A lack of understanding of the philosophical underpinnings of energy conditions makes it difficult to determine which formulation is correct.  Rather than try to answer that question in general, we note that, because we are using energy conditions in lieu of the equation of state, it is reasonable to suppose that the physical formulation is appropriate for our purposes.

The energy conditions therefore take the standard form for a perfect fluid
\begin{subequations}
\begin{eqnarray}
\hbox{null:} &&\quad \rho+p\geq0\; , \\
\hbox{weak:} &&\quad \rho\geq0\; \hbox{and}\; \rho+p\geq0\; , \\
\hbox{strong:} &&\quad \rho+3p\geq0\; , \label{strong}\\
\hbox{dominant:} &&\quad \rho\geq|p|\;
\end{eqnarray}
\end{subequations}
Because we have already assumed positive energy density, the weak and null conditions are equivalent for our purposes.  The strong energy condition is derived in standard gravity by insisting that gravity is attractive and using Einstein's equations to derive Eq.~(\ref{strong}).  Since Einstein's equations do not apply in non-standard gravity, we have no particular reason to demand that this form of the strong energy condition apply. The dominant energy condition is derived from the premise that mass-energy does not propagate faster than light.  This is a reasonable expectation for matter, so we will accept it.  In addition, the dominant energy condition implies the weak and null energy conditions, so we don't have to consider them separately.  We will therefore take the dominant energy condition to supply our cutoff point.  Since, for all these energy conditions, the constraints on $p$ are proportional to $\rho$, we divide through by $\rho$ to get constraints on $p/\rho$ that are constant and dimensionless.

 \begin{figure}[h]
	\centering
	\subfloat{\includegraphics[angle=0,width=3.1in,clip]{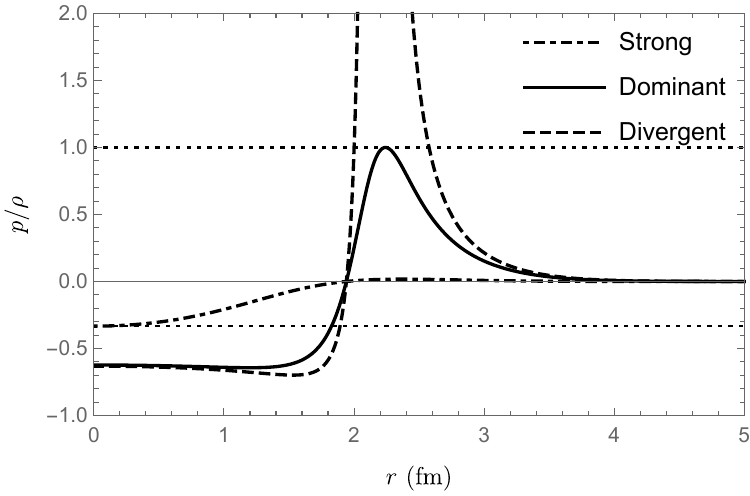}}\hfill
	\caption{The pressure to density ratio for the Gaussian density profile as a function of radius corresponding to the cutoff values of $\lambda$ as given in Table~\ref{table:cutoffs}.}
	\label{fig:Gaussian}
\end{figure}

\begin{figure}[h]
	\centering
	\subfloat{\includegraphics[angle=0,width=3.1in,clip]{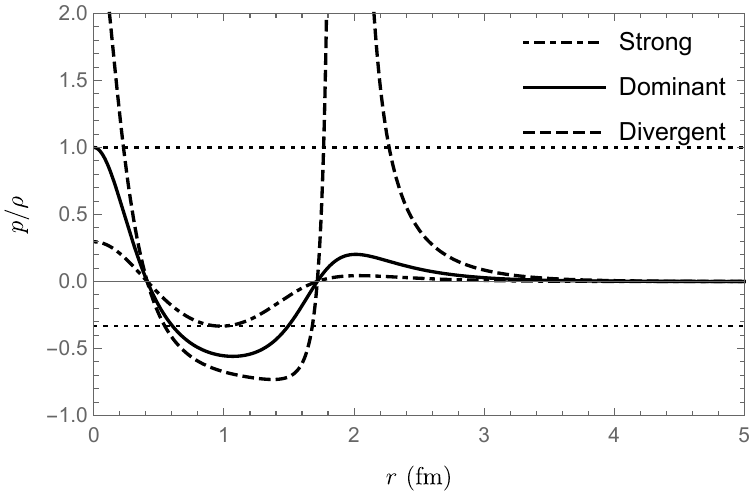}}\hfill
	\caption{The pressure to density ratio for the experimental density profile as a function of radius corresponding to the cutoff values of $\lambda$ as given in Table~\ref{table:cutoffs}.}
	\label{fig:Experimental}
\end{figure}

\begin{table}[h]
	\centering
	\begin{tabular}{| c | c | c | c | c |}
		\hline
		{$\lambda~(\upmu\hbox{m}^2)$} & Gaussian & Experimental  \\ \hline
		Strong & 2.46 & 2.72 \\ \hline
		Dominant  & 9.35 & 5.00 \\ \hline
		Divergent & 10.54 & 8.24 \\  \hline
	\end{tabular}	
	\caption{Limiting values of $\lambda$ for the Gaussian  and experimental density profiles at which the strong energy condition is first violated, the dominant condition is violated, and the value for which the pressure diverges.}
	\label{table:cutoffs}
\end{table}

In the case of the Gaussian, Eq.~(\ref{P}) can be solved analytically to give
\begin{equation}\label{Gaussian p}
\frac{p}{\rho} = \frac{24\pi G\frac{\lambda^2\rho_0}{\sigma^2}e^{-\frac{r^2}{2\sigma^2}}\!\left(\frac{r^2}{\sigma^2}-4\right)}
	{1 - 48\pi G\frac{\lambda^2\rho_0}{\sigma^2}e^{-\frac{r^2}{2\sigma^2}}\!\left(\frac{r^2}{\sigma^2}-3\right)}\; .
\end{equation}
This is plotted for some interesting values of $\lambda$ in Fig.~\ref{fig:Gaussian}.  The ``experimental" distribution does not seem to lend itself to analytical solution.  It is plotted in Fig.~\ref{fig:Experimental}.  The strong and dominant energy inequalities are saturated for the values of $\lambda$ listed in Table~\ref{table:cutoffs}.  Note that for a finite value of $\lambda$, the denominator appearing in Eq.~(\ref{P}) vanishes, causing the pressure to diverge at some radius, so the density profiles we used are impossible to maintain for any equation of state.  This value is also included in Table~\ref{table:cutoffs}.  At this point, the assumption $|p| \lesssim \rho$ becomes invalid, along with the approximations derived from it, $8\pi G |\lambda p| \ll 1$ and $|\lambda \rho^{\prime\prime}| \gg |p|$, but these will still be valid until the ratio $\frac{p}{\rho}$ is implausibly large.

It should be noted that the choice of the Gaussian density distribution causes a change in the limits by less than a factor of two, so we see that our limit is not strongly shape dependent. Because we believe the experimental density profile is realistic and the dominant energy condition must be satisfied, we will treat $| \lambda | < 5\times 10^{-12}\,\hbox{m}^2$ as our limit.   This is well within our region of applicability, $10^{-30}\,{\rm m^2} \ll | \lambda | \ll 10^8\,{\rm m}^2$.

~

\section{Conclusions}\label{Conclusions}
We have studied alternate theories of gravity with non-minimal coupling between matter and curvature.  We argued that in a large class of such theories the nuclear realm is an excellent place to look for experimental limits.  Because our argument is based on the effect of density gradients, systems which merely have high density, like neutron stars, would provide much weaker limits.

We illustrated our approach using a theory where the matter Lagrangian is multiplied by a factor of $1+ \lambda R$. Using the $^4$He nucleus we found a limit $|\lambda| < 5 \times 10^{-12}\,{\rm m}^2$, more than thirty orders of magnitude stronger than the best previous limit we know of \cite{Bertolami_2008b}.  We anticipate that our approach will produce strong limits in any model with a non-trivial matter coupling $f_2(R)$.  Theories with only $f_1(R)$, such as $f(R)$ chameleon theories \cite{PhysRevD.78.104021}, will likely not be so strongly constrained.

It should be noted that in the parameter range we focused on, the perturbative contributions proportional to $\lambda$ are negligible, but there are large effects coming from $\lambda^2$ contributions.  Much of the previous work kept only linear terms in $\lambda$ \cite{Bertolami_2013, Castel_Branco_2014,March_2019}, and therefore missed what may be the largest contributions to new physics from non-minimally coupled gravity.

\begin{acknowledgments}
We would like to thank T.~Ordines for his assistance in preparing this document.
\end{acknowledgments}

\input{limitsv7.bbl}
\end{document}

%% file: limitsv7.bbl
%